\documentclass[conference]{IEEEtran}
\IEEEoverridecommandlockouts

\usepackage{balance}
\usepackage[T1]{fontenc}
\usepackage{ifthen}
\usepackage[cmex10]{amsmath} % Use the [cmex10] option to ensure
\usepackage[lined,boxed,commentsnumbered,linesnumbered, ruled]{algorithm2e}
\usepackage[url,hyperrefblack,notheorems,IEEEtran]{research17}
\usepackage{lipsum} % print dummy text
 \usepackage{comment}
\usepackage{booktabs} % required for showing good \hline in matrix form --> \midrule
\usepackage{mathtools}
\usepackage{amsmath,amssymb,amsfonts}
\usepackage{tikz} % common graphical tools for pictures
\usetikzlibrary{shapes, patterns,decorations.text, decorations.pathreplacing}
\newcommand{\tikzmark}[1]{\tikz[overlay,remember picture] \node (#1){};}
\newtheorem{lemma}{\mylemmaname}
\newtheorem{theorem}{\mytheoremname}
\newtheorem{definition}{\mydefinitionname}

\usepackage[capitalize]{cleveref}
\crefname{equation}{\unskip}{\unskip}
\crefname{claim}{Claim}{Claims} %\crefname{type}{singular}{plural}
\usepackage{array}
\usepackage{multirow}
\newcolumntype{C}[1]{>{\centering\arraybackslash}p{#1}}
\allowdisplaybreaks

\renewcommand{\vect}[1]{\vectg{#1}} % switch by default to second version!
\renewcommand{\vmat}[1]{\bm{\mat{#1}}} % notation for random matrices
\newcommand{\code}[1]{\mathscr{#1}} % the code notation: seen as a collection
\newcommand{\collect}[1]{\mathscr{#1}} % the collection notation
\newcommand*{\Resize}[2][4]{\resizebox{#1}{!}{\ensuremath{#2}}} % Resize based on line or column width
\makeatletter
\renewcommand*\env@matrix[1][*\c@MaxMatrixCols c]{%
  \hskip -\arraycolsep
  \let\@ifnextchar\new@ifnextchar
  \array{#1}}
\makeatother
\newcommand{\HP}[1]{\HH\left(#1\right)} 
\newcommand{\eHP}[1]{\HH(#1)} 
\newcommand{\bigHP}[1]{\HH\bigl(#1\bigr)}
\newcommand{\BigHP}[1]{\HH\Bigl(#1\Bigr)}
\newcommand{\HPcond}[2]{\HH\left(#1 \kern0.1em\middle|\kern0.1em #2\right)}
\newcommand{\eHPcond}[2]{\HH(#1 \kern0.1em|\kern0.1em #2)} 
\newcommand{\bigHPcond}[2]{\HH\bigl(#1 \kern-0.1em \bigm| \kern-0.1em#2\bigr)}
\newcommand{\BigHPcond}[2]{\HH\Bigl(#1 \kern-0.1em \Bigm| \kern-0.1em#2\Bigr)}
\newcommand{\MI}[2]{\II\left(#1 \kern0.1em{;}\kern0.1em #2\right)} 
\newcommand{\eMI}[2]{\II(#1 \kern0.1em{;}\kern0.1em #2)} 
\newcommand{\bigMI}[2]{\II\bigl(#1 \kern0.1em{;}\kern0.1em #2\bigr)}
\newcommand{\BigMI}[2]{\II\Bigl(#1 \kern0.1em{;}\kern0.1em #2\Bigr)}
\newcommand{\MIcond}[3]{\II\left(#1 \kern0.1em{;}\kern0.1em #2 \kern0.1em\middle|\kern0.1em #3\right)}
\newcommand{\eMIcond}[3]{\II(#1 \kern0.1em{;}\kern0.1em #2 \kern0.1em|\kern0.1em #3)} 
\newcommand{\bigMIcond}[3]{\II\bigl(#1 \kern0.1em{;}\kern0.1em #2 \kern-0.1em \bigm| \kern-0.1em#3\bigr)}
\newcommand{\BigMIcond}[3]{\II\Bigl(#1 \kern0.1em{;}\kern0.1em #2 \kern-0.1em \Bigm| \kern-0.1em#3\Bigr)}
\renewcommand{\r}{\color{red}} % red
\renewcommand{\b}{\color{blue}} % blue
\DeclareSymbolFont{matha}{OML}{txmi}{m}{it}% txfonts
\DeclareMathSymbol{\varv}{\mathord}{matha}{118}
\begin{document}
%
% paper title
% Titles are generally capitalized except for words such as a, an, and, as,
% at, but, by, for, in, nor, of, on, or, the, to and up, which are usually
% not capitalized unless they are the first or last word of the title.
% Linebreaks \\ can be used within to get better formatting as desired.
% Do not put math or special symbols in the title.
\sloppy \title{Capacity of Private Linear Computation for Coded Databases \thanks{This work is supported by US NSF grant
    CNS-1526547 and the Research Council of Norway (grant 240985/F20).}}

% author names and affiliations
% use a multiple column layout for up to three different
% affiliations
\author{\IEEEauthorblockN{Sarah A.~Obead\IEEEauthorrefmark{2}, Hsuan-Yin Lin\IEEEauthorrefmark{3}, Eirik
    Rosnes\IEEEauthorrefmark{3}, and J{\"o}rg Kliewer\IEEEauthorrefmark{2}
    \IEEEauthorblockA{\IEEEauthorrefmark{2}Helen and John C.~Hartmann
      Department of Electrical and Computer Engineering \\ New Jersey Institute of Technology, Newark, New Jersey 07102, USA}
    \IEEEauthorblockA{\IEEEauthorrefmark{3}Simula UiB, N--5020 Bergen, Norway}}}

% make the title area
\maketitle

% As a general rule, do not put math, special symbols or citations
% in the abstract
\begin{abstract}
  We consider the problem of private linear computation (PLC) in a distributed storage system. In PLC, a user wishes to
  compute a linear combination of $f$ messages stored in noncolluding databases while revealing no information about the
  coefficients of the desired linear combination to the databases. In extension of our previous work we employ 
  linear codes to encode the information on the databases. We show that the PLC capacity, which is the ratio of the
  desired linear function size and the total amount of downloaded information, matches the maximum distance separable
  (MDS) coded capacity of private information retrieval for a large class of linear codes that includes MDS codes. In
  particular, the proposed converse is valid for any number of messages and linear combinations, and the capacity
  expression depends on the rank of the coefficient matrix obtained from all linear combinations.
  % Finally, we also provide examples of linear code families which are able
  % to achieve the PLC capacity.
\end{abstract}

% no keywords
\section{Introduction}
\label{sec:introduction}

Private information retrieval (PIR) from public databases has been the focus of attention for several decades in the
computer science community (see, e.g., \cite{ChorKushilevitzGoldreichSudan98_1,Yekhanin10_1}). Recently, the aspect of
minimizing the communication cost, e.g., the required rate or bandwidth of privately querying the databases with the
desired requests and downloading the corresponding information from the databases has in particular been of interest
also in the information theory and coding communities (see, e.g.,
\cite{ShahRashmiRamchandran14_1,ChanHoYamamoto15_1,TajeddineGnilkeElRouayheb18_1app,SunJafar17_1,SunJafar18_2,BanawanUlukus18_1,
  FreijHollantiGnilkeHollantiKarpuk17_1,KumarRosnesGraellAmat17_1}). Those settings primarily focus on distributed
storage systems (DSSs), where data is encoded by a linear code and then stored across several databases.

A recently proposed generalization of the PIR problem
\cite{MirmohseniMaddahAli17_1sub,SunJafar17_1sub,Karpuk18_1,ObeadKliewer18_1,ChenWangJafar18_1} addresses the
\emph{private computation (PC)} of functions of these messages. %also denoted as private function retrieval.
In PC a user has access to a given number of databases and intends to compute a function of messages stored in these
databases. This function is kept private from the databases, as they may be under the control of an adversary. In
\cite{MirmohseniMaddahAli17_1sub,SunJafar17_1sub}, the authors consider that the data is replicated at each database,
which are assumed to be noncolluding. Both works characterize the fundamental information-theoretic communication
overhead needed to privately compute a given \emph{linear} function, called private linear computation (PLC), and
specify the corresponding capacity and achievable rates as a function of the number of messages and the number of
databases, respectively. Interestingly, the obtained PLC capacity is equal to the PIR capacity of
\cite{SunJafar17_1}. The extension to the coded case is addressed in \cite{Karpuk18_1,ObeadKliewer18_1}. In
\cite{Karpuk18_1}, a PC scheme for polynomial functions of the data over $t$ colluding databases is constructed by
generalizing the star-product PIR scheme of \cite{FreijHollantiGnilkeHollantiKarpuk17_1}. Specifically, for
systematically coded storage, this scheme considers functions that are polynomials of fixed degree $g$, and achieves a
rate matching the best asymptotic rate (when the number of messages goes to infinity) of coded PIR
$\const{R}=1-\gamma_{\mathsf{c}}$ for $g=t=1$ (the case of linear function retrieval and noncolluding databases), where
$\gamma_{\mathsf{c}}$ denotes the code rate. In our previous work \cite{ObeadKliewer18_1}, we consider maximum distance
separable (MDS) coded storage and the computation of linear message functions. The presented scheme is able to achieve
the MDS-coded PIR capacity in \cite{BanawanUlukus18_1}, referred to as the MDS-PIR capacity in the sequel,  and strictly
generalizes the achievable schemes in \cite{MirmohseniMaddahAli17_1sub,SunJafar17_1sub}.

On the other hand, for the case of noncolluding databases, a PIR protocol for a DSS where data is stored using a non-MDS
linear code, is proposed in \cite{KumarRosnesGraellAmat17_1}. The protocol is shown to achieve the asymptotic MDS-PIR
capacity for several example codes. In \cite{KumarLinRosnesGraellAmat17_1sub,LinKumarRosnesGraellAmat18_1}, this
approach is extended to colluding databases and it is proved that both the asymptotic and the nonasymptotic MDS-PIR
capacity can be achieved for a large class of linear codes in the noncolluding case. For noncolluding databases, the
first family of non-MDS codes for which the PIR capacity is known is found in
\cite{LinKumarRosnesGraellAmat18_2app,LinKumarRosnesGraellAmat18_3sub}. Further, PIR on linearly-coded databases for the case of colluding databases is also
addressed in \cite{FreijHollantiGnilkeHollantiKarpuk17_1,TajeddineGnilkeElRouayheb18_1app,FreijHollantiGnilkeHollanti17_1sub}. For the PLC case however,
capacity results for arbitrary linear codes have not been addressed so far in the open literature to the best of our
knowledge.

In this work, we intend to fill this void and propose an extension of our previous works in \cite{ObeadKliewer18_1} and
\cite{LinKumarRosnesGraellAmat18_1} to PLC for arbitrary linear codes with noncolluding databases. Surprisingly, we show
that the PLC capacity for a large class of linear codes matches the MDS-PIR capacity when a user wishes to compute an
arbitrary linear combination of $f$ independent equal-sized messages over some finite field $\Field_q$, distributed over
$n$ noncolluding linearly-coded databases. In particular, by extending our previous work \cite{ObeadKliewer18_1} where
only achievable rates are stated, we also adapt the converse proof of \cite[Thm.~4]{LinKumarRosnesGraellAmat18_3sub}
to the PLC problem in this paper. In contrast to \cite{SunJafar17_1sub}, our converse is valid for any number of
messages $f$ and any number of desired linear combinations $\mu$, and our capacity result depends on the rank of the
coefficient matrix obtained from all $\mu$ linear combinations.
% Finally, we also provide examples of classes of codes which are able to achieve the MDS-PIR capacity. Our results
% demonstrate that, compared to the na\"{i}ve scheme, where $f$ coded messages are downloaded and linearly combined
% offline at the user (requiring $f$-times the coded PIR rate), downloading the result of the computation privately and
% directly from the databases does not incur any penalty in rate compared to the coded PIR case.

\section{Definitions and Problem Statement}
\label{sec:definitions-problem}

\subsection{Notation and Definitions}
\label{sec:notation-definitions}

We denote by $\Naturals$ the set of all positive integers and for some $a,b\in\Naturals$, $a \leq b$, 
$[a:b]\eqdef\{a,a+1,\ldots,b\}$. Random and deterministic quantities are carefully distinguished as follows. A random
variable is denoted by a capital Roman letter, e.g., $X$, while its realization is denoted by the corresponding small
Roman letter, e.g., $x$; vectors are boldfaced, e.g., $\vect{X}$ denotes a random vector and $\vect{x}$ denotes a
deterministic vector; random matrices are represented by bold sans serif letters, e.g., $\vmat{X}$, and $\mat{X}$
represents its realization. In addition, sets are denoted by calligraphic upper case letters, e.g., $\set{X}$, and
$\comp{\set{X}}$ denotes the complement of a set $\set{X}$ in a universe set. For a given index set $\set{S}$, we also
write $\vmat{X}^\set{S}$ and $Y_\set{S}$ to represent $\bigl\{\vmat{X}^{(v)}\colon v\in\set{S}\bigr\}$ and
$\bigl\{Y_j\colon j\in\set{S}\bigr\}$, respectively. Furthermore, some constants are also depicted by Greek letters or a
special font, e.g., $\const{X}$. We denote a submatrix of $\mat{X}$ that is restricted in columns by the set $\set{I}$
by $\mat{X}|_{\set{I}}$. The function $\HP{\cdot}$ represents the entropy of its argument and $\MI{X}{Y}$ denotes the
mutual information of the random variables $X$, $Y$. $\trans{(\cdot)}$ denotes the transpose of its argument, while
$\rank{\mat{V}}$ denotes the rank of a matrix $\mat{V}$. We use the customary code parameters $[n,k]$ to denote a code
$\code{C}$ over the finite field $\Field_q$ of blocklength $n$ and dimension $k$. A generator matrix of $\code{C}$ is
denoted by $\mat{G}^{\code{C}}$. % while $\code{C}^{\mat{G}}$ represents the corresponding linear code generated by
%$\mat{G}$. 
A set of coordinates of $\code{C}$, $\set{I}\subseteq[1:n]$, of size $k$ is said to be an \emph{information
  set} if and only if $\mat{G}^\code{C}|_\set{I}$ is invertible.  The function $\chi(\vect{x})$ denotes the support of a
vector $\vect{x}$.
% while the support of a linear code $\code{C}$ is defined as the set of coordinates where not all codewords are
% zero. The $s$-th generalized Hamming weight of an $[n,k]$ code $\code{C}$, denoted by $d_s^{\code{C}}$, $s\in [1:k]$,
% is defined as the cardinality of the smallest support of an $s$-dimensional subcode of $\code{C}$.

\subsection{System Model}
\label{sec:system-model}
% \subsection{Private Linear Computation for Coded DSSs}
% \label{sec:coded-PLC}

The PLC problem for coded DSSs is described as follows. We consider a DSS that stores $f$ messages
$\vmat{W}^{(1)},\ldots,\vmat{W}^{(f)}$, where each message $\vmat{W}^{(m)}=\bigl(W_{i,j}^{(m)}\bigr)$, $m\in [1:f]$, can
be seen as a random matrix of size $\beta\times k$ over $\Field_q$ with $\beta,k \in\Naturals$. Each message is encoded
using an $[n,k]$ code as follows. Let $\vect{W}^{(m)}_i=\bigl(W^{(m)}_{i,1},\ldots,W^{(m)}_{i,k}\bigr)$,
$i\in [1:\beta]$, be a message vector corresponding to the $i$-th row of $\vmat{W}^{(m)}$. Each $\vect{W}^{(m)}_i$ is
encoded by an $[n,k]$ code $\code{C}$ over $\Field_q$ into a length-$n$ codeword
$\vect{C}^{(m)}_i=\bigl(C^{(m)}_{i,1},\ldots,C^{(m)}_{i,n}\bigr)$. The $\beta f$ generated codewords $\vect{C}_i^{(m)}$
are then arranged in the array $\vmat{C}=\trans{\bigl(\trans{(\vmat{C}^{(1)})}|\ldots|\trans{(\vmat{C}^{(f)})}\bigr)}$
of dimensions $\beta f \times n$, where
$\vmat{C}^{(m)}=\trans{\bigl(\trans{(\vect{C}^{(m)}_1)}|\ldots|\trans{(\vect{C}^{(m)}_{\beta})}\bigr)}$. The code
symbols $C_{1,j}^{(m)},\ldots,C_{\beta,j}^{(m)}$, $m\in[1:f]$, for all $f$ messages are stored on the $j$-th database,
$j\in[1:n]$.

Let $\const{L}\eqdef \beta\cdot k$. Then, each message $\vmat{W}^{(m)}$, $m\in[1:f]$, can also be seen as a random
vector variable $\vmat{W}^{(m)}=(W^{(m)}_1,\ldots,W^{(m)}_\const{L})$ of $\const{L}$ symbols that are chosen
independently and uniformly at random from $\Field_q$. Thus,
\begin{IEEEeqnarray*}{rCl}
  \HP{\vmat{W}^{(m)}}& = &\const{L},\,\forall \,m\in[1:f],
  \\
  \HP{\vmat{W}^{(1)},...,\vmat{W}^{(f)}}& = &f\const{L}\quad (\textnormal{in } q\textnormal{-ary units}).
\end{IEEEeqnarray*}

\subsection{Private Linear Computation for Coded DSSs}
\label{sec:private-linear-computation_coded-DSSs}

We consider the case where no set of databases can collude.  A user wishes to privately compute exactly one
linear function from the $\mu$ \emph{candidate} linear functions $\vmat{X}^{(1)},\ldots,\vmat{X}^{(\mu)}$ from the
coded DSS, where $\vmat{X}^{(v)}=\bigl(X^{(v)}_1,\ldots,X^{(v)}_\const{L}\bigr)$. The $\mu$-tuple
$\trans{\bigl(X^{(1)}_l,\ldots,X^{(\mu)}_l\bigr)}$, $\forall\,l\in [1:\const{L}]$, is mapped by a
deterministic matrix $\mat{V}$ of size $\mu\times f$ over $\Field_q$ by
\begin{IEEEeqnarray}{rCl}
  \begin{pmatrix}
    X^{(1)}_l
    \\
    \vdots
    \\
    X^{(\mu)}_l
  \end{pmatrix}=\mat{V}_{\mu\times f}
  \begin{pmatrix}
    W^{(1)}_l
    \\
    \vdots
    \\
    W^{(f)}_l
  \end{pmatrix}.\label{eq:linear-mappingV}
\end{IEEEeqnarray}
Hence, the user privately generates an index $v\in[1:\mu]$ and wishes to compute the $v$-th linear function while
keeping the index $v$ private from each database. Here, we also assume that the rank of $\mat{V}$ is equal to
$\rank{\mat{V}}=r\leq\min\{\mu,f\}$ and the indices corresponding to a basis for the row space of $\mat{V}$ are denoted
by the set $\set{L}\triangleq\{\ell_1,\ldots,\ell_r\}\subseteq [1:\mu]$.

In order to retrieve the linear function $\vmat{X}^{(v)}$ from the coded DSS, $v\in[1:\mu]$, the user sends a query
$Q^{(v)}_j$ over $\Field_q$ to the $j$-th database for all $j\in[1:n]$. Since the queries are generated by the user
without any prior knowledge of the realizations of the candidate linear functions, the queries are independent of the
linear functions. In other words, we have
\begin{IEEEeqnarray*}{c}
  \MI{\vmat{X}^{(1)},\ldots,\vmat{X}^{(\mu)}}{Q^{(v)}_1,\ldots,Q^{(v)}_n}=0,\,\forall\,v\in[1:\mu].
\end{IEEEeqnarray*}
In response to the received query, database $j$ sends the answer $A^{(v)}_j$ back to the user. $A^{(v)}_j$ is a
deterministic function of $Q^{(v)}_j$ and the data stored in the database. Thus, $\forall\,v\in[1:\mu]$, 
\begin{IEEEeqnarray*}{rCl}
  \BigHPcond{A^{(v)}_j}{Q^{(v)}_j,\vmat{C}^{[1:f]}}=0,\,\forall\,j\in[1:n].
\end{IEEEeqnarray*}

Next, we define a PLC protocol for coded DSSs as follows.
\begin{definition}
  \label{Def:perfect-PLC}
  Consider a DSS with $n$ noncolluding databases storing $f$ messages using an $[n,k]$ code. The user wishes to
  retrieve the $v$-th linear function $\vmat{X}^{(v)}$,  $v\in[1:\mu]$, from the available information $Q^{(v)}_j$ and
  $A^{(v)}_j$, $j\in[1:n]$. For a PLC protocol, the following conditions must be satisfied
  $\forall\,v\in[1:\mu]$,
  \begin{IEEEeqnarray*}{rCl}    
    \IEEEeqnarraymulticol{3}{l}{%
      \text{[Privacy]} }\nonumber\\*\qquad%
    && \bigMI{v}{Q^{(v)}_j,A^{(v)}_j,\vmat{X}^{(1)},\ldots,\vmat{X}^{(\mu)}}=0,\,\forall\,j\in[1:n],
    %\IEEEyesnumber\IEEEyessubnumber\IEEEeqnarraynumspace %\label{eq:privacy}
    \\
    \IEEEeqnarraymulticol{3}{l}{%
      \text{[Recovery]} }\nonumber\\*\qquad%
    && \bigHPcond{\vmat{X}^{(v)}}{A^{(v)}_1,\ldots,A^{(v)}_n,Q^{(v)}_1,\ldots,Q^{(v)}_n}=0.
    %\IEEEyessubnumber %\label{eq:recovery}
  \end{IEEEeqnarray*}
\end{definition}

To measure the efficiency of a PLC protocol, we consider the required number of downloaded symbols for retrieving the
$\const{L}$ symbols of the candidate linear function.
\begin{definition}[PLC Rate and Capacity for Coded DSSs]
  \label{def:def_PLCrate}
  The rate of a PLC protocol, denoted by $\const{R}$, is the ratio of the desired linear function size $\const{L}$ to
  the total required download cost $\const{D}$, i.e.,
  \begin{IEEEeqnarray*}{c}
    \const{R}\eqdef\frac{\const{L}}{\const{D}}.
  \end{IEEEeqnarray*}
  The PLC \emph{capacity} $\const{C}_\textnormal{PLC}$ is the maximum possible PLC rate over all possible PLC protocols
  for a given $[n,k]$ storage code.
\end{definition}

\subsection{MDS-PIR Capacity-Achieving Codes}
\label{sec:MDS-PIRcapacity-achieving-codes_PLC}

In \cite{KumarLinRosnesGraellAmat17_1sub,LinKumarRosnesGraellAmat18_1}, a PIR protocol for any linearly-coded DSS that
uses an $[n,k]$ code to store $f$ messages, named Protocol~1, is proposed. The corresponding PIR rate depends on the
following property of the underlying storage code $\code{C}$.
\begin{definition}
  \label{def:PIRachievable-rate-matrix}
  Let $\code{C}$ be an arbitrary $[n,k]$ code. A $\nu\times n$ binary matrix $\mat{\Lambda}_{\kappa,\nu}(\code{C})$ is
  said to be a \emph{PIR achievable rate matrix} for $\code{C}$ if the following conditions are satisfied.
  \begin{enumerate}
  \item \label{item:1} The Hamming weight of each column of $\mat{\Lambda}_{\kappa,\nu}$ is $\kappa$, and
  \item \label{item:2} for each matrix row $\vect{\lambda}_i$, $i\in[1:\nu]$, $\chi(\vect{\lambda}_i)$ always contains
    an information set.
  \end{enumerate}
  In other words, each coordinate $j$ of $\code{C}$, $j\in [1:n]$, appears exactly $\kappa$ times in
  $\{\chi(\vect{\lambda}_i)\}_{i\in [1:\nu]}$, and every set $\chi(\vect{\lambda}_i)$ contains an information set.
\end{definition}

In \cite{KumarLinRosnesGraellAmat17_1sub,LinKumarRosnesGraellAmat18_1}, it is shown that the MDS-PIR capacity (i.e.,
the PIR capacity for a DSS where data is encoded and stored using an MDS code) can be achieved using Protocol~1 for a
special class of $[n,k]$ codes. In particular, to achieve the MDS-PIR capacity using Protocol~1, the $[n,k]$ storage
code should possess a specific underlying structure as given by the following theorem.

% The following lemma gives the underlying structure of an $[n,k]$ code that achieves the MDS-PIR capacity is.
\begin{theorem}[\cite{KumarLinRosnesGraellAmat17_1sub,LinKumarRosnesGraellAmat18_1}]
  \label{lem:MDS-PIRcapacity-achieving-matrix}
  Consider a DSS that uses an $[n,k]$ code $\code{C}$ to store $f$ messages. If a PIR achievable rate matrix
  $\mat{\Lambda}_{\kappa,\nu}(\code{C})$ with $\frac{\kappa}{\nu}=\frac{k}{n}$ exists, then the MDS-PIR capacity
  \begin{IEEEeqnarray*}{rCl}
    \const{C}_{\textnormal{MDS-PIR}}\eqdef\Bigl(1-\frac{k}{n}\Bigr)\inv{\left[1-\Bigl(\frac{k}{n}\Bigr)^f\right]}
    \label{eq:PIRcapacity}  
  \end{IEEEeqnarray*}
  is achievable.
\end{theorem}

This gives rise to the following definition.
\begin{definition}
  \label{def:MDS-PIRcapacity-achieving-codes}
  A PIR achievable rate matrix $\mat{\Lambda}_{\kappa,\nu}(\code{C})$ with $\frac{\kappa}{\nu}=\frac{k}{n}$ for an
  $[n,k]$ code $\code{C}$ is called an \emph{MDS-PIR capacity-achieving} matrix, and $\code{C}$ is referred to as an
  \emph{MDS-PIR capacity-achieving} code.
\end{definition}

In Section~\ref{sec:achievable-scheme_coded-PLC}, we will present a PLC protocol and a general achievable rate for any $\rank{\mat{V}}=r$ by using
the PIR achievable rate matrix $\mat{\Lambda}_{\kappa,\nu}$ of an $[n,k]$ code.

\section{PLC Capacity for MDS-PIR Capacity-Achieving Codes}
\label{sec:PLCcapacity_MDS-PIRcapacity-achieving-codes}

The main result of this paper is the derivation of the PLC capacity for a coded DSS where data is encoded and stored using
a linear code from the class of MDS-PIR capacity-achieving codes introduced in
\cite{KumarLinRosnesGraellAmat17_1sub,LinKumarRosnesGraellAmat18_1},  which is stated in the following theorem.
\begin{theorem}
  \label{thm:converse_MDS-PLCcapacity-achieving-codes}
  Consider a DSS that uses an $[n,k]$ MDS-PIR capacity-achieving code $\code{C}$ to store $f$ messages. Then, the
  maximum achievable PLC rate over all possible PLC protocols, i.e., the PLC capacity, is 
  \begin{IEEEeqnarray}{c}
    \const{C}_{\textnormal{PLC}}\eqdef\Bigl(1-\frac{k}{n}\Bigr)\inv{\left[1-\Bigl(\frac{k}{n}\Bigr)^r\right]}
    \label{eq:PLCcapacity}
  \end{IEEEeqnarray}
  for any rank $r$ of the linear mapping from \eqref{eq:linear-mappingV}.
\end{theorem}

The achievable scheme is provided in Section~\ref{sec:achievable-scheme_coded-PLC}, while the converse proof is given in
Section~\ref{sec:converse-proof_coded-PLC}. Note that the class of MDS-PIR capacity-achieving codes includes MDS codes, cyclic
codes, Reed-Muller codes, and certain classes of distance-optimal local reconstruction codes
\cite{KumarLinRosnesGraellAmat17_1sub,LinKumarRosnesGraellAmat18_1}. 

We remark here that it is known that if $\rank{\mat{V}}=f$, then the PLC capacity for an MDS-coded DSS is equal to the
MDS-PIR capacity \cite{ObeadKliewer18_1}, i.e., if $\rank{\mat{V}}=f$, then
\begin{IEEEeqnarray*}{rCl}
  \const{C}_{\textnormal{PLC}}=\Bigl(1-\frac{k}{n}\Bigr)\inv{\left[1-\Bigl(\frac{k}{n}\Bigr)^f\right]}=
  \const{C}_{\textnormal{MDS-PIR}}.
\end{IEEEeqnarray*}

\section{Achievable Scheme}
\label{sec:achievable-scheme_coded-PLC}

We start with a coded PIR query scheme for a message of $\mu$ \emph{dependent} virtual messages. Given that the messages
are stored using an $[n,k]$ MDS-PIR capacity-achieving code $\code{C}$, we find the associated $\nu\times n$ MDS-PIR
capacity-achieving matrix $\mat{\Lambda}_{\kappa,\nu}$ and obtain the PIR interference matrices
$\mat{A}_{\kappa\times n}$ and $\mat{B}_{(\nu-\kappa)\times n}$ as given by the following definition.

\begin{definition}[\cite{KumarLinRosnesGraellAmat17_1sub,LinKumarRosnesGraellAmat18_1}]
  \label{def:PIRinterference-matrices}
  For a given $\nu\times n$ PIR achievable rate matrix $\mat{\Lambda}_{\kappa,\nu}(\code{C})=(\lambda_{u,j})$, we define
  the PIR interference matrices $\mat{A}_{\kappa{\times}n}=(a_{i,j})$ and $\mat{B}_{(\nu-\kappa){\times}n}=(b_{i,j})$
  for the code $\code{C}$ with
  \begin{IEEEeqnarray*}{rCl}
    a_{i,j}& \eqdef &u \text{ if } \lambda_{u,j}=1,\,\forall j \in [1:n], i \in[1:\kappa], u \in  [1:\nu],\\
    b_{i,j}& \eqdef &u \text{ if } \lambda_{u,j}=0,\,\forall j \in [1:n], i \in[1:\nu-\kappa], u \in  [1:\nu].
  \end{IEEEeqnarray*}
\end{definition}

Note that in \cref{def:PIRinterference-matrices}, for each $j\in [1:n]$, distinct values of $u\in [1:\nu]$ should be
assigned for all $i$. Thus, the assignment is not unique in the sense that the order of the entries of each column of
$\mat{A}$ and $\mat{B}$ can be permuted. 
% For $j\in [1:n]$, let $\set{A}_j\eqdef\{a_{i,j}\colon i\in [1:\kappa]\}$ and
% $\set{B}_j\eqdef\{b_{i,j}\colon i\in [1:\nu-\kappa]\}$. Note that the $j$-th column of $\mat{A}$ contains the row
% indices of $\mat{\Lambda}$ whose entries in the $j$-th column are equal to $1$, while $\mat{B}$ contains the remaining
% row indices of $\mat{\Lambda}$. Hence, it can be observed that $\set{B}_j=[1:\nu]\setminus\set{A}_j$,
% $\forall\,j\in [1:n]$.
Next, for the sake of illustrating our query set generation algorithm, we make use of the
following definition.
\begin{definition}
  \label{def:uCoordinateSet_A}
  By $\set{S}(u|\mat{A}_{\kappa \times n})$ we denote the set of column coordinates of matrix
  $\mat{A}_{\kappa{\times}n}=(a_{i,j})$ in which at least one of its entries is equal to $u$, i.e.,
  \begin{IEEEeqnarray*}{rCl}
    \set{S}(u|\mat{A}_{\kappa\times n})\eqdef\{j\in [1:n]\colon\exists\,a_{i,j}=u,i\in [1:\kappa]\}.
  \end{IEEEeqnarray*}
\end{definition}

As a result, we require the size of the message to be $\const{L}= \nu^{\mu}\cdot k $ (i.e.,~${\beta=\nu^\mu}$). The
scheme is completed in $\mu$ rounds.
% which are repeated $\kappa$ times

\subsection{Query Generation}
\label{sec:query-generation}

Before running the main algorithm to generate query sets, the following index preparation for the coded
symbols stored in each database is performed.

%Before we start the query sets generation in the main
%Algorithm~\ref{alg:generation_QuerySet}: \texttt{Q-Gen}, we first perform the following index preparation for the coded
%symbols stored in each database.

\indent {\bf \textit{1) Index Preparation}:} The goal is to make the coded symbols queried from each database to appear
to be chosen randomly and independently from the desired linear function
index. Note that the linear function is computed separately for the $t$-th
row of all messages, $t \in [1:\beta]$. Therefore,  similar to the PLC scheme
in \cite{SunJafar17_1sub} and the MDS-coded PLC scheme in
\cite{ObeadKliewer18_1}, we apply a permutation that is fixed across all
coded symbols for the $t$-th row to maintain the
dependency across the associated message elements. Let $\pi(\cdot)$ be a random permutation function over
$[1:\beta]$, and let
\begin{IEEEeqnarray*}{c}
  U^{(v)}_{t,j}\triangleq \vect{v}_{v}{\vect{C}}_{\pi(t),j},\quad t\in[1:\beta], \quad j\in[1:n],
\end{IEEEeqnarray*}
denote the $t$-th permuted code symbol associated with the $v$-th virtual message ${\vmat{X}^{(v)}}$ stored in the $j$-th
database. Here, $\vect{v}_{v}$ represents the $v$-th row vector of the matrix $\mat{V}_{\mu\times f}=(v_{i,j})$ and 
$\vect{C}_{t,j}\eqdef\trans{\bigl(C_{t,j}^{(1)},\ldots,C_{t,j}^{(f)}\bigr)}$.
% , and the random variable $\sigma_t$ is used to indicate the sign assigned to each individual coded symbol,
% $\sigma_t \in \{+1,-1\}$ \cite[Sec.~4.2]{SunJafar17_1sub}. Both $\sigma_t$ and
The permutation $\pi(\cdot)$ is randomly selected privately and uniformly by the user.

\indent {\bf \textit{2) Preliminaries}:}
For $\tau\in[1:\mu]$, a sum $U^{(v_1)}_{i_1,j} + U^{(v_2)}_{i_2,j} + \cdots + U^{(v_\tau)}_{i_\tau,j}$, $j\in[1:n]$, of
$\tau$ distinct symbols is a $\tau$-sum for any subset $\{i_1,i_2,\ldots,i_{\tau}\}\subseteq [1:\beta]$, and
$\{v_1,v_2,\ldots,v_{\tau}\}\subseteq [1:\mu]$ determines the \emph{type} of the $\tau$-sum.
% a $\tau$-sum is defined a sum of $\tau$ distinct elements out of $\mu$ elements.
Since we have $\binom{\mu}{\tau}$ different selections of $\tau$ distinct elements out of
$\mu$ elements, a $\tau$-sum can have $\binom{\mu}{\tau}$ different types.
% we denote each different selection as the \emph{type} of the $\tau$-sum. {\m In other words,
%
%
The query generation procedure is subdivided into $\mu$ rounds. 
%
%
%
%
%
% For % A query $Q^{(v)}_j$, $j\in[1:n]$, sent to the $j$-th database consists of a set of queries in a total of $\mu$ rounds,
% where a round represents a group of all $\binom{\mu}{\tau}$ types of $\tau$-sums for all $\tau\in[1:\mu]$, resulting
% in $\mu$ rounds in total.
%
For a requested linear function indexed by $v\in[1:\mu]$, a query set $Q^{(v)}_j$, $j\in[1:n]$, is composed of $\mu$
disjoint subsets, one generated by each round $\tau\in[1:\mu]$. For each round the query subset is further subdivided
into two subsets. The first subset $Q^{(v)}_j(\set{D};\tau)$ consists of $\tau$-sums with a single symbol from the
  \emph{desired} linear function and $\tau-1$ symbols from \emph{undesired} linear functions, while the second
subset $Q^{(v)}_j(\set{U};\tau)$ contains $\tau$-sums with symbols only from undesired linear functions.
% Note that we require $\kappa^{\mu-(\tau-1)}(\nu-\kappa)^{\tau-1}$ distinct instances of each $\tau$-sum type for every
% query set $Q_j^{(v)}(\set{D};\tau)$.
%
% Note that in the case of coded databases using an MDS-PIR capacity-achieving code, we will use $\kappa$ information
% sets associated with the matrix $\mat{A}$ to recover the requested linear function. Moreover, the \emph{side
% information}, rows from undesired functions forming $\tau$-sum types with symbols from the \emph{desired} function
% acting as interference to maintain the privacy of the desired function, are generated based on the matrix $\mat{B}$
% and decoded by utilizing the code coordinates forming an information set in the code array.
%
To this end, in the scheme we introduce $\kappa n$ auxiliary query sets $Q^{(v)}_j(a_{i,j},\set{D};\tau)$,
$i\in[1:\kappa]$, where each query consists of a distinct symbol from the desired linear function and $\tau-1$ symbols from undesired linear functions,
and $(\nu-\kappa) n$ auxiliary query sets $Q^{(v)}_j(b_{i,j},\set{U};\tau)$, $i\in[1:\nu-\kappa]$, to represent the side
information query sets for each database $j\in [1:n]$. We utilize these sets to generate the query sets of each round
according to the PIR interference matrices $\mat{A}_{\kappa\times n}$ and $\mat{B}_{(\nu-\kappa)\times n}$.
The query sets for all databases are generated by Algorithm~\ref{alg:generation_QuerySet} through the following
procedures.\footnote{Note that a query $Q_j^{(v)}$ sent to the $j$-th database usually indicates the row indices of
  the stored code symbols that the user requests, while the answer $A_j^{(v)}$ to the query $Q_j^{(v)}$ refers to the
  particular code symbols requested through the query. In Algorithm~\ref{alg:generation_QuerySet}, with some abuse of
  notation, the generated queries are sets containing their answers.}

\begin{algorithm}[htbp]
  \caption{\texttt{Q-Gen}}
  \label{alg:generation_QuerySet}
  \SetInd{0.35em}{0.35em}
  \small
  \SetKwFunction{InitialRound}{Initial-Round}
  \SetKwFunction{Partition}{Partition}
  \SetKwFunction{DesiredQ}{Desired-Q}
  \SetKwFunction{ExploitSI}{Exploit-SI}
  \SetKwFunction{SetAddition}{SetAddition}
  \SetKwFunction{MSym}{M-Sym}  
  \SetKwInOut{Input}{Input}
  \SetKwInOut{Output}{Output}
  \SetKwComment{Comment}{$\triangleright$\ }{}{}
  \DontPrintSemicolon
  
  \Input{$v$, $\mu$, $\kappa$, $\nu$, $n$, $\mat{A}_{\kappa\times n}$, and $\mat{B}_{(\nu-\kappa)\times n}$}
  \Output{${Q}^{(v)}_1,\ldots,{Q}^{(v)}_n$}
  %$Q^{(v)}_j\leftarrow\emptyset$, $j\in[1:n]$\;
  \For{$\tau\in[1:\mu]$}{    
  $Q^{(v)}_j(\mathcal{D};\tau) \leftarrow\emptyset$, $Q^{(v)}_j(\mathcal{U};\tau) \leftarrow\emptyset$,  $j\in[1:n]$\;
    $\alpha_{\tau}\leftarrow\kappa^{\mu-1}+\sum_{h=1}^{\tau-1}\binom{\mu-1}{h}\kappa^{\mu-(h+1)}(\nu-\kappa)^{h}$\;
    \Comment{\b Generate query sets for the initial round}
    \If{$\tau=1$}{
      \For{$u\in [1:\nu]$}{
        \For{$j\in \set{S}(u|\mat{A}_{\kappa\times n})$}{
          $Q^{(v)}_j(u,\set{D};\tau),Q^{(v)}_j(u,\set{U};\tau)\leftarrow\InitialRound(u,\alpha_\tau,j,v,\tau)$\;     
        }      
      }
    }
    \Comment{\b Generate query sets for the following rounds $\tau>1$}
    \Else{
      \For{$u\in [1:\nu]$}{
        \Comment{\b Generate desired symbols for the following rounds $\tau>1$}
        \For{$j\in \set{S}(u|\mat{A}_{\kappa\times n})$}{
          $Q^{(v)}_j(u,\set{D};\tau)\leftarrow\DesiredQ(u,\alpha_\tau,j,v,\tau)$\;}
        \Comment{\b Generate side information for the following rounds $\tau>1$}
        \For{$j\in\set{S}(u|\mat{B}_{(\nu-\kappa)\times n})$}{
          $Q^{(v)}_j(u,\set{U};\tau-1)\leftarrow
          \ExploitSI(u,Q^{(v)}_1(u,\mathcal{U},\tau-1),\ldots,Q^{(v)}_n(u,\mathcal{U},\tau-1),j,v,\tau)$\;}
      }
      \Comment{\b Generate the final desired query sets for the following rounds $\tau>1$}
        \For{$j\in [1:n]$}{
          $\tilde{Q}^{(v)}_j(\set{U};\tau-1)\leftarrow\bigcup\limits_{i\in [1:\nu-\kappa]}Q^{(v)}_j(b_{i,j},\set{U};\tau-1)$\;
          $\tilde{Q}^{(v)}_j(1,\set{U};\tau-1),\ldots, \tilde{Q}^{(v)}_j(\kappa,\set{U};\tau-1) \leftarrow
          \Partition\bigl(\tilde{Q}^{(v)}_j(\set{U};\tau-1)\bigr)$\;
          \For{$i\in [1:\kappa]$}{
            $Q^{(v)}_j(a_{i,j},\set{D};\tau)\leftarrow
            \SetAddition\bigl(Q_j^{(v)}(a_{i,j},\set{D};\tau),\tilde{Q}^{(v)}_j(i,\set{U};\tau-1)\bigr)$\;          
          }      
        }
      \Comment{\b Generate the query sets of undesired symbols by forcing message symmetry for the following rounds $\tau>1$}  
      \For{$u\in [1:\nu]$}{
        \For{$j\in\set{S}(u|\mat{A}_{\kappa\times n})$}{
          $Q^{(v)}_j(u,\set{U};\tau)\leftarrow\MSym\bigl(Q_j^{(v)}(u,\set{D};\tau),j,v,\tau\bigr)$\;}
      }
    }
      \For{$u\in [1:\nu]$}{
        \For{$j\in \set{S}(u|\mat{A}_{\kappa\times n})$}{
          $Q^{(v)}_j(\set{D};\tau)\leftarrow Q^{(v)}_j(\set{D};\tau) \cup Q^{(v)}_j(u,\set{D};\tau)$\;
          $Q^{(v)}_j(\set{U};\tau)\leftarrow Q^{(v)}_j(\set{U};\tau) \cup Q^{(v)}_j(u,\set{U};\tau)$\;
        }
      }
  }  
  \For{$j\in [1:n]$}{
    $Q^{(v)}_j\leftarrow\bigcup\limits_{\tau\in [1:\mu]}\Bigl(Q^{(v)}_j(\set{D};\tau)\cup 
    Q^{(v)}_j(\set{U};\tau)\Bigr)$\;
  }
\end{algorithm}

\indent {\bf \textit{3) Initialization (Round ${\tau=1}$)}:} In the initialization step, the algorithm generates the
auxiliary queries for the first round. This round is described by Steps {5} to {11} of
Algorithm~\ref{alg:generation_QuerySet}, where we have $\tau=1$ for the $\tau$-sum. At this point,
Algorithm~\ref{alg:generation_QuerySet} invokes the subroutine \texttt{Initial-Round} given in
Algorithm~\ref{alg:initial-round} to generate $Q^{(v)}_j(a_{i,j},\set{D};1)$, $i\in[1:\kappa]$, such that each of these
query sets contains $\alpha_1=\kappa^{\mu-1}$ distinct code symbols. Furthermore, to maintain function symmetry, the
algorithm asks each database for the same number of distinct symbols of all other linear functions in
$Q^{(v)}_j(a_{i,j},\set{U};1)$, $i\in[1:\kappa]$, resulting in a total number of $\binom{\mu-1}{1}\kappa^{\mu-1}$
symbols. As a result, the queried code symbols in the auxiliary query sets for each database are symmetric with respect
to all linear function vectors indexed by $v'\in [1:\mu]$. We associate the symbols of undesired functions in $\kappa$
groups, each placed in the undesired query sets $Q^{(v)}_j(a_{i,j},\set{U};1)$, $i\in[1:\kappa]$, to be exploited as
distinct side information for the second round of the scheme. Since this procedure produces $\kappa$ undesired query
sets for each database, database symmetry is maintained.

\begin{algorithm}[htbp]
  \caption{\texttt{Initial-Round}}
  \label{alg:initial-round}
  \SetInd{0.35em}{0.35em}
  \small
  \SetKwFunction{new}{new}
  \SetKwInOut{Input}{Input}
  \SetKwInOut{Output}{Output}
  \SetKwComment{Comment}{$\triangleright$\ }{}
  \DontPrintSemicolon	 
  
  \Input{$u$, $\alpha_\tau$, $j$, $v$, and $\tau$}
  \Output{$\phi^{(v)}(u,\set{D};\tau), \phi^{(v)}(u,\set{U};\tau)$}
  
  $\phi^{(v)}(u,\set{D};\tau)\leftarrow\emptyset$, $\phi^{(v)}(u,\set{U};\tau)\leftarrow\emptyset$\;
  % \Comment{\textcolor{blue}{Generate query set for the initial round}}
  \For{$l\in [1:\alpha_{\tau}]$}{
    $\phi^{(v)}(u,\set{D};\tau)\leftarrow\phi^{(v)}(u,\set{D};\tau)\cup\bigl\{U^{(v)}_{({\r u}-1)
      \cdot\alpha_\tau+l,j}\bigr\}$\;
    $\phi^{(v)}(u,\set{U};\tau)\leftarrow\phi^{(v)}(u,\set{U};\tau)\ \cup$
    $\phantom{abcdefghijklmn}
    \left(\bigcup\limits_{v'=1}^{\mu}\bigl\{U^{(v')}_{({\r u}-1)\cdot\alpha_\tau+l,j}\bigr\}
      \setminus\bigl\{U^{(v)}_{({\r u}-1)\cdot\alpha_\tau+l,j}\bigr\}\right)$\;
  }
\end{algorithm}

\indent {\bf \textit{4) Desired Function Symbols for Rounds ${\tau>1}$}:} For the following rounds a similar process is
repeated in terms of generating auxiliary query sets containing distinct code symbols from the desired linear
function $\vmat{U}^{(v)}=(U^{(v)}_{t,j})$. This is accomplished in Steps {16} to {18} by calling the subroutine
$\texttt{Desired-Q}$, given in Algorithm~\ref{alg:Desired-setQ_following-rounds}, to generate
$Q^{(v)}_j(a_{i,j},\set{D};\tau)$, $i\in[1:\kappa]$, such that each of these query sets contains
$(\alpha_{\tau}-1)-\alpha_{\tau-1}+1=\binom{\mu-1}{\tau-1}\kappa^{\mu-(\tau-1+1)}(\nu-\kappa)^{\tau-1}$ distinct code
symbols from the desired linear function $\vmat{U}^{(v)}$.

\begin{algorithm}[htbp]
  \caption{\texttt{Desired-Q}}
  \label{alg:Desired-setQ_following-rounds}
  \SetInd{0.35em}{0.35em}
  \small
  \SetKwInOut{Input}{Input}
  \SetKwInOut{Output}{Output}
  \SetKwComment{Comment}{$\triangleright$\ }{}
  \DontPrintSemicolon	   
  \Input{$u$, $\alpha_\tau$, $j$, $v$, and $\tau$}
  \Output{$\phi^{(v)}(u,\set{D};\tau)$}  
  $\phi^{(v)}(u,\set{D};\tau)\leftarrow\emptyset$\;
  % \Comment{\b Generate the query set of desired symbols w.r.t.\ a particular $u$ for the following rounds $\tau>1$}
  \For{$l\in[\alpha_{\tau-1}:\alpha_\tau-1]$}{
    $\phi^{(v)}(u,\set{D};\tau)\leftarrow\phi^{(v)}(u,\set{D};\tau)
    \cup\bigl\{U^{(v)}_{l\cdot\nu+{\r u},j}\bigr\}$\;
  }
\end{algorithm}

\indent {\bf \textit{5) Side Information Exploitation}:} In Steps {20} to {22}, we generate the side information query sets $Q^{(v)}_j(b_{i',j},\set{U};\tau-1)$, $i'\in [1:\nu-\kappa]$, from the auxiliary query sets $Q^{(v)}_1(a_{i,1},\set{U};\tau-1),\ldots,Q^{(v)}_n(a_{i,n},\set{U};\tau-1)$, $i\in [1:\kappa]$, of the previous round $\tau-1$, $\tau\in[2:\mu]$, by applying the subrountine \texttt{Exploit-SI}, given by
Algorithm~\ref{alg:Exploit-SI}. This subroutine is extended from \cite{SunJafar17_1sub} based on our coded storage
scenario.  Note
that in Algorithm~\ref{alg:Exploit-SI} the function \texttt{Reproduce}$(j,Q^{(v)}_{j'}(u,\set{U};\tau-1))$,
$j'\in [1:n]\setminus\{j\}$, simply reproduces all the queries in the auxiliary query set
$Q^{(v)}_{j'}(u,\set{U};\tau-1)$ with a different coordinate $j$.

% we generate the query sets for the rounds {$\tau\in[1:\mu-1]$}

\begin{algorithm}[htbp]
  \caption{\texttt{Exploit-SI}}
  \label{alg:Exploit-SI}
  \SetInd{0.35em}{0.35em}
  \small
  \SetKwFunction{Reproduce}{Reproduce}
  \SetKwInOut{Input}{Input}
  \SetKwInOut{Output}{Output}
  \SetKwComment{Comment}{$\triangleright$\ }{}
  \DontPrintSemicolon	 
  
  \Input{$u$, $Q^{(v)}_1(u,\set{U};\tau-1),\ldots,Q^{(v)}_n(u,\set{U};\tau-1)$, $j$, $v$, and $\tau$} %, %$j\in [1:n]\setminus\{j\}$}
  \Output{$\phi^{(v)}(u,\set{U};\tau-1)$}
  
  $\phi^{(v)}(u,\set{U};\tau-1)\leftarrow\emptyset$\;  
  \For{$i\in [1:\kappa]$}{
    \For{$j'\in[1:n]\setminus\{j\}$}{
      \If{$u=a_{i,j'}$}{
        $\phi^{(v)}(u,\set{U};\tau-1)\leftarrow\Reproduce(j,Q^{(v)}_{j'}(u,\set{U};\tau-1))$\;
        break\;
      }
    }
  }
\end{algorithm}

Next, we update the desired query sets $Q^{(v)}_j(a_{i,j},\set{D};\tau)$ in Steps {25} to {31}. First, the function
$\texttt{Partition}\bigl(\tilde{Q}^{(v)}_j(\set{U};\tau-1)\bigr)$ denotes a procedure that divides a set into $\kappa$
disjoint equally-sized subsets. This is viable since based on the subroutine \texttt{Initial-Round} and the
  following subroutine \texttt{M-Sym}, one can show that
  $\bigcard{\tilde{Q}^{(v)}_j(\set{U};\tau-1)}
  =\binom{\mu-1}{\tau-1}\kappa^{\mu-(\tau-1)}(\nu-\kappa)^{(\tau-1)-1}\cdot(\nu-\kappa)$ for each round
  $\tau\in[2:\mu]$, which is always divisible by $\kappa$. Secondly, we assign the new query set of
desired symbols $Q^{(v)}_j(a_{i,j},\set{D};\tau)$ for the current round by using an element-wise set addition
$\texttt{SetAddition}(Q_1,Q_2)$. The element-wise set addition is defined as
$\bigl\{q_{i_l}+q_{i'_l}\colon q_{i_l}\in Q_1,q_{i'_l}\in Q_2, l\in [1:\rho]\bigr\}$ with $\card{Q_1}=\card{Q_2}=\rho$,
where $\rho$ is an appropriate integer.  In Steps {33} to {37}, the subroutine $\texttt{M-Sym}$, given in
Algorithm~\ref{alg:message-symmetry}, is invoked to generate the undesired query sets $Q^{(v)}_j(a_{i,j},\set{U};\tau)$
by utilizing message symmetry. This subroutine selects symbols of undesired messages to generate $\tau$-sums that
enforce symmetry in the round queries. The procedure resembles the subroutine \texttt{M-Sym} proposed in
\cite{SunJafar17_1sub}.
% For completeness, we describe the detailed steps of
In Algorithm~\ref{alg:message-symmetry},
% according to our coded storage scenario, where
$\Pi_\tau$ denotes the set of all possible selections of $\tau$ distinct indices in $[1:\mu]$ and
$\texttt{Lexico}(\Pi_\tau)$ denotes the corresponding set of ordered selections (the indices $(v_1,\ldots,v_\tau)$ of a
selection of $\Pi_\tau$ are ordered in natural lexicographical order).
% In Algorithm~\ref{alg:message-symmetry},
Further, the notation $U^{(v_x)}_{\ast,j}$ implies that the row index of the code symbol can be arbitrary. This is the
case since only the linear function indices $(v_1,\ldots,v_\tau)$ are necessary to determine $i_z$,
$\forall\,z\in [1:\tau]$. As a result, symmetry over the linear functions is maintained.
%
% for each type of $\tau$-sum query, we have $\kappa^{\mu-(\tau-1)}(\nu-\kappa)^{\tau-1}$ query sets for each round.
%
% Note that we require of each $\tau$-sum type for every query set $Q_j^{(v)}(\set{D};\tau)$.

\begin{algorithm}[htbp!]
  \caption{\texttt{M-Sym}}
  \label{alg:message-symmetry}
  \SetInd{0.35em}{0.35em}
  \small
  \SetKwFunction{Msym}{M-sym}
  \SetKwFunction{Lexico}{Lexico}
  \SetKwInOut{Input}{Input}
  \SetKwInOut{Output}{Output}
  \SetKwComment{Comment}{$\triangleright$\ }{}
  \DontPrintSemicolon	 
  
  \Input{$Q^{(v)}_j(u,\set{D};\tau)$, $j$, $v$, and $\tau$}
  \Output{$\phi^{(v)}(u,\set{U};\tau)$}
  
  $\phi^{(v)}(u,\set{U};\tau)\leftarrow\emptyset$\;
  \For{$(v_1,\ldots,v_\tau)\in\Lexico(\Pi_\tau)$, $v\notin \{v_1,\ldots,v_\tau \}$}{
    $\phi^{(v)}(u,\set{U};\tau)\leftarrow\phi^{(v)}(u,\set{U};\tau)\cup
    \bigl\{U^{(v_1)}_{i_1,j}+\ldots+U^{(v_\tau)}_{i_\tau,j}\bigr\}$ $\textnormal{ such that }
    \forall\,z\in[1:\tau]$, $\exists\,U^{(v)}_{i_z,j}+\sum\limits_{\substack{x\in [1:\tau]\\ x\neq z}}
    U^{(v_x)}_{\ast,j}\in Q^{(v)}_j(u,\set{D};\tau)$\;
  }
\end{algorithm}

\indent {\bf \textit{6) Query Set Assembly}:} Finally, in Steps {39} to {48}, we assemble each query set from
disjoint query subsets obtained in all $\tau$ rounds. It can be shown that
  $Q_j^{(v)}(\set{D};\tau) \cup Q_j^{(v)}(\set{U};\tau)$ contains $\kappa^{\mu-(\tau-1)}(\nu-\kappa)^{\tau-1}$
  $\tau$-sums for every $\tau$-sum type. % distinct instances

% \vspace{-2ex}
\subsection{Sign Assignment and Redundancy Elimination}
\label{sec:SignAssyment}
%\vspace{-0.5ex}

After Algorithm~\ref{alg:generation_QuerySet}, the user will know which row indices of the stored code symbols
  he/she is going to request. To reduce the total number of downloaded symbols, the linear dependency among the candidate
  linear functions is exploited. We adopt a similar sign assignment process over $\sigma^{(v)}_t\in \{+1,-1\}$ to each
symbol in the query sets: $\sigma^{(v)}_t U^{(v)}_{t,j}$, based on the desired linear function index $v$ as
introduced in \cite[Sec.~4.2]{SunJafar17_1sub}. Here, the sign $\sigma^{(v)}_t$ is privately generated by the user with a uniform
  distribution over $\{-1,+1\}$. The intuition behind the sign assignment is to introduce a uniquely solvable equation
system from the different $\tau$-sum types given the side information available from all other databases. By obtaining
such a system of equations in each round, the user can determine some of the queries offline to decode the desired
linear function and/or interference, thus reducing the download rate. Based on this insight we can state the following
lemma.

%\begin{lemma}[\hspace{-0.1ex}\cite{SunJafar17_1sub}] \label{lem:redundancy} 
\begin{lemma}
  \label{lem:redundancy}
  For all {$v\in[1:\mu]$}, each database $j\in[1:n]$, and based on the side information available from the databases,
  there are $\mu-r\choose \tau$ redundant $\tau$-sum types out of all possible types $\mu\choose \tau$ in each round
  {${\tau\in[1:\mu-r]}$} of the query sets.
\end{lemma}

Since repetition codes and MDS codes are MDS-PIR capacity-achieving codes, Lemma~\ref{lem:redundancy} generalizes
\cite[Lem.~1]{SunJafar17_1sub} and \cite[Lem.~1]{ObeadKliewer18_1}.
% Lemma~\ref{lem:redundancy} is also applicable here as in \cite{ObeadKliewer18_1} and \cite{SunJafar17_1sub}.
We can extend the proof of 
\cite[Lem.~1]{SunJafar17_1sub}  to our scheme based on the insight that the redundancy  resulting from the linear
dependencies between messages is also present for MDS-PIR capacity-achieving codes.
% the desired linear function is performed over linearly-coded databases as for the MDS-coded databases. Hence, the
% redundancy resulting from the linear dependencies between messages is also present for MDS-PIR capacity-achieving
% codes and we can extend the proof of \cite[Lem.~1]{SunJafar17_1sub} to our scheme.
We now make the final modification to our PLC query sets. We first directly apply the sign assignment $\sigma^{(v)}_t$,
then remove all $\tau$-sums corresponding to redundant $\tau$-sum types from every round $\tau\in{[1:\mu]}$ (see
Lemma~\ref{lem:redundancy}). Note that, in our case the redundancy is dependent on the rank of the functions matrix,
$\rank{\mat{V}}=r\leq \min\{\mu,f\}$, thus generalizing the MDS-coded PLC case. Finally, we generate the queries
$Q^{(v)}_{1:n}$.

\subsection{Example} \label{ex:PLCex_n4k2f2mu3}

To illustrate the key idea of the scheme we use the following example.

%\begin{example}
  %\label{ex:PLCex_n4k2f2mu3}
  Consider four messages $\vmat{W}^{(1)}$, $\vmat{W}^{(2)}$, $\vmat{W}^{(3)}$, and $\vmat{W}^{(4)}$ that are stored in a
  DSS using a $[4,2]$ MDS-PIR capacity-achieving code $\code{C}$ with generator matrix
	
  \begin{IEEEeqnarray*}{c}
    \mat{G}^{\code{C}}=
    \begin{pmatrix}
      1 & 0 &1 & 1\\ 
      0 & 1 &1 & 1  
    \end{pmatrix}
    \text{ for which }
  %\end{IEEEeqnarray*}
  %
  %One can easily verify that 	
  %\begin{IEEEeqnarray*}{rCl}
    \mat{\Lambda}_{1,2}=
    \begin{pmatrix}
      1 & 0 &1 & 0 \\
      0 & 1 &0 & 1
    \end{pmatrix}
  \end{IEEEeqnarray*}  
  is a PIR achievable rate matrix. According to Definition~\ref{def:PIRinterference-matrices} we obtain the PIR
  interference matrices %\vspace{-0.5ex}
  %\begin{IEEEeqnarray*}{c}
    $\mat{A}_{1\times 4}=
    \begin{pmatrix} 1 & 2 &1 & 2 \end{pmatrix}$ and 
    $\mat{B}_{1\times 4} =\begin{pmatrix} 2 & 1 &2 & 1 \end{pmatrix}$. %\vspace{-0.5ex} 
  %\end{IEEEeqnarray*}
  Suppose that the user wishes to obtain a linear function $\vmat{X}^{(v)}$ from a set of $\mu=3$ candidate linear
  functions, i.e.,~$v\in[1:3]$, defined by
  % such that the deterministic matrix $\mat{V}\in {\mathbb F}_3^{\mu\times f}$ is given by \vspace{-0.5ex}
  \begin{IEEEeqnarray*}{rCl}
    \mat{V}_{\mu\times f} = \begin{pmatrix}
      1 & 0 &0 & 1 \\ 
      1 & 1 &0 & 0 \\
      2 & 1 &0 & 1  
    \end{pmatrix}.%\vspace{-0.5ex} 
  \end{IEEEeqnarray*}
  
  We simplify notation by letting $x_{t,j}=U^{(1)}_{t,j}$, $y_{t,j}=U^{(2)}_{t,j}$, and $z_{t,j}=U^{(3)}_{t,j}$ for all
  ${t\in[1:\beta]}$, $j\in[1:n]$, where $\beta=\nu^\mu=8.$ Let the desired linear function index be $v=1$. For this
  example, the construction of the query sets is briefly presented in the following steps.
  
  \indent {\bf \textit{Initialization (Round ${\tau=1}$)}:} Algorithm~\ref{alg:generation_QuerySet} starts with
  ${\tau=1}$ to generate auxiliary query sets for each database holding $\kappa^{\mu-1}=1$ distinct instances of
  $x_{t,j}$. By message symmetry this also applies to $y_{t,j}$ and $z_{t,j}$. The auxiliary query sets for the first
  round are given in Table~\ref{tab:axQ-table}(a). Note that the queries for $z_{t,j}$ can be generated offline by the
  user and thus are later removed from the query sets.

  \indent {\bf \textit{Following Rounds (${\tau>1}$)}:} As can be seen from Table~I(b) and (c), using the interference
  matrices $\mat{A}_{1\times 4}$ and $\mat{B}_{1\times 4}$, the algorithm further generates the auxiliary query sets
  $Q^{(1)}_{j}(a_{1,j},\set{D};\tau)$ that contain new symbols to be queried jointly with symbols of the desired linear
  function during the exploitation of side information for the $j$-th database, $j\in [1:n]$. After generating the
  desired auxiliary query sets $Q^{(1)}_j(a_{1,j},\set{D};\tau)$, the undesired auxiliary query sets $Q^{(1)}_j(a_{1,j},\set{U};\tau)$ are
  generated by enforcing message symmetry. Finally, we apply the sign assignment scheme to the query sets and remove
  redundant queries from each database. The resulting queries are shown in Table~\ref{tab:answers-table}.

  The PLC rate of the achievable scheme is  equal to
  $\frac{\nu^\mu\cdot k}{\const{D}}=\frac{8\cdot 2}{6\cdot 4}=\frac{2}{3}$, which is equal to the PLC capacity in
  \eqref{eq:PLCcapacity} with $\rank{\mat{V}}=2$.

  \begin{table}[htbp!]  
  \centering
  \caption{Auxiliary query sets for each round. The magenta dashed arrows and the cyan arrows indicate that the
    \texttt{Exploit-SI} algorithm and the \texttt{M-Sym} algorithm are used, respectively.}
  \label{tab:axQ-table}
  \vskip -1mm 
  \Resize[\columnwidth]{
    \begin{IEEEeqnarraybox}[
      \IEEEeqnarraystrutmode
      \IEEEeqnarraystrutsizeadd{3pt}{2pt}]{v/c/v/c/v/c/v/c/v/c/v}
      \IEEEeqnarrayrulerow\\
      & j && 1  && 2 && 3 && 4\\
      \hline\hline
      & Q^{(1)}_j(a_{1,j},\set{D};1)
      && x_{({\r 1}-1)\cdot 1+1,1} &&  x_{({\r 2}-1)\cdot 1+1,2} && x_{({\r 1}-1)\cdot 1+1,3} && x_{({\r 2}-1)\cdot 1+1,4} &
      \\*\hline
      & \multirow{2}{*}{$Q^{(1)}_j(a_{1,j},\set{U};1)$}
      && y_{({\r 1}-1)\cdot 1+1,1}\tikzmark{y11} &&  y_{({\r 2}-1)\cdot 1+1,2} && y_{({\r 1}-1)\cdot 1+1,3}\tikzmark{y13}
      && y_{({\r 2}-1)\cdot 1+1,4} &
      \\
      & && z_{({\r 1}-1)\cdot 1+1,1}\tikzmark{z11} &&  z_{({\r 2}-1)\cdot 1+1,2} && z_{({\r 1}-1)\cdot 1+1,3}\tikzmark{z13}
      && z_{({\r 2}-1)\cdot 1+1,4} &
      \\*\IEEEeqnarrayrulerow
    \end{IEEEeqnarraybox}}
  (a)
  \\[2mm]
  \Resize[\columnwidth]{
    \begin{IEEEeqnarraybox}[
      \IEEEeqnarraystrutmode
      \IEEEeqnarraystrutsizeadd{3pt}{2pt}]{v/c/v/c/v/c/v/c/v/c/v}
      \IEEEeqnarrayrulerow\\
      & j && 1  && 2 && 3 && 4\\
      \hline\hline
      & \multirow{2}{*}{$Q^{(1)}_j(a_{1,j},\set{D};2)$}
      && x_{1\cdot 2+{\r 1},1}+y_{2,1} &&  x_{1\cdot 2+{\r 2},2}+y_{1,2}\tikzmark{y22} && x_{1\cdot 2+{\r 1},3}+y_{2,3}
      && x_{1\cdot 2+{\r 2},4}+y_{1,4}\tikzmark{y24} &
      \\
      & && x_{2\cdot 2+{\r 1},1}+z_{2,1} && x_{2\cdot 2+{\r 2},2}+z_{1,2}\tikzmark{z22} && x_{2\cdot 2+{\r 1},3}+z_{2,3}
      && x_{2\cdot 2+{\r 2},4}+z_{1,4}\tikzmark{z24} &
      \\*\hline
      & Q^{(1)}_j(a_{1,j},\set{U};2)
      && y_{4+{\r 1},1}+z_{2+{\r 1},1} && y_{4+{\r 2},2}+z_{2+{\r 2},2}\tikzmark{yz22}
      && y_{4+{\r 1},3}+z_{2+{\r 1},3} && y_{4+{\r 2},4}+z_{2+{\r 2},4}\tikzmark{yz24} &
      \\*\IEEEeqnarrayrulerow
    \end{IEEEeqnarraybox}}
  (b)
  \\[2mm]
  \Resize[\columnwidth]{
    \begin{IEEEeqnarraybox}[
      \IEEEeqnarraystrutmode
      \IEEEeqnarraystrutsizeadd{3pt}{2pt}]{v/c/v/c/v/c/v/c/v/c/v}
      \IEEEeqnarrayrulerow\\
      & j && 1  && 2 && 3 && 4\\
      \hline\hline
      & Q^{(1)}_j(a_{1,j},\set{D};3)
      && x_{3\cdot 2+{\r 1},1}+y_{6,1}+z_{4,1}\tikzmark{xyz11} && x_{3\cdot 2+{\r 2},2}+y_{5,2}+z_{3,2}
      && x_{3\cdot 2+{\r 1},3}+y_{6,3}+z_{4,3}\tikzmark{xyz13} && x_{3\cdot 2+{\r 2},4}+y_{5,4}+z_{3,4} &
      \\*\IEEEeqnarrayrulerow
    \end{IEEEeqnarraybox}}
  (c)
  \tikz[overlay,remember picture]{
    \draw[magenta,dashed,>=stealth,->,thick] ($(y11)+(-1.0,-0.65)$) to ($(y22)+(-2.0,-0.1)$);
    \draw[magenta,dashed,>=stealth,->,thick] ($(y13)+(-1.65,-0.65)$) to ($(y22)+(-2.0,-0.1)$);
    \draw[magenta,dashed,>=stealth,->,thick] ($(y11)+(-1.0,-0.65)$) to ($(y24)+(-3.1,-0.1)$);
    \draw[magenta,dashed,>=stealth,->,thick] ($(y13)+(-1.65,-0.65)$) to ($(y24)+(-3.1,-0.1)$);
    \path[cyan,>=stealth,->,thick] (z22)+(-1.65,0) edge [bend left] ($(yz22)+(-1.75,0)$);
    \path[cyan,>=stealth,->,thick] (z24)+(-2.9,0) edge [bend left] ($(yz24)+(-2.95,0)$);
    \draw[magenta,dashed,>=stealth,->,thick] ($(yz22)+(-2.5,-0.15)$) to ($(xyz11)+(-2.6,0.1)$);
    \draw[magenta,dashed,>=stealth,->,thick] ($(yz22)+(-2.5,-0.15)$) to ($(xyz13)+(-5.2,0.1)$);
    \draw[magenta,dashed,>=stealth,->,thick] ($(yz24)+(-3.65,-0.15)$) to ($(xyz11)+(-2.6,0.1)$);
    \draw[magenta,dashed,>=stealth,->,thick] ($(yz24)+(-3.65,-0.15)$) to ($(xyz13)+(-5.2,0.1)$);
  }
  \vskip -2mm 
\end{table}

  \begin{table}[htbp!]
  \centering
  \caption{Final query sets (after sign assignment and removal of redundant queries) for rounds one to three for a coded DSS that uses the $[4,2]$ code
    of Section~\ref{ex:PLCex_n4k2f2mu3} to store $f=4$ messages.}
  \label{tab:answers-table}
  \vskip -1mm
  \Resize[\columnwidth]{
    \begin{IEEEeqnarraybox}[
      \IEEEeqnarraystrutmode
      \IEEEeqnarraystrutsizeadd{3pt}{2pt}]{v/c/v/c/v/c/v/c/v/c/v}
      \IEEEeqnarrayrulerow\\
      & j && 1 && 2 && 3 && 4\\
      \hline\hline
      & Q^{(1)}_j(\set{D};1)
      && x_{1,1} &&  x_{2,2} && x_{1,3} && x_{2,4} &
      \\*\cline{1-11}      
      & Q^{(1)}_j(\set{U};1)
      && y_{1,1} &&  y_{2,2} && y_{1,3} && y_{2,4} &
      \\*\cline{1-11}      
      & \multirow{2}{*}{$Q^{(1)}_j(\set{D};2)$}
      && x_{3,1}-y_{2,1} &&  x_{4,2}-y_{1,2} && x_{3,3}-y_{2,3} && x_{4,4}-y_{1,4} &
      \\ %*\cline{3-11}
      & 
      && x_{5,1}-z_{2,1} &&  x_{6,2}-z_{1,2} && x_{5,3}-z_{2,3} && x_{6,4}-z_{1,4} &
      \\*\cline{1-11}      
      & Q^{(1)}_j(\set{U};2)
      && y_{5,1}-z_{3,1} &&  y_{6,2}-z_{4,2} && y_{5,3}-z_{3,3} && y_{6,4}-z_{4,4} &
      \\*\cline{1-11}      
      & Q^{(1)}_j(\set{D};3)
      && x_{7,1}-y_{6,1}+z_{4,1} && x_{8,2}-y_{5,2}+z_{3,2}
      && x_{7,3}-y_{6,3}+z_{4,3} && x_{8,4}-y_{5,4}+z_{3,4} &
      \\*\IEEEeqnarrayrulerow
    \end{IEEEeqnarraybox}}
    %\vskip -2mm 
\end{table}

  %\vspace{10ex}
%\end{example}

\subsection{Achievable Rate}
% The proof of optimality follows from the structure of the queries and
% Lemma~\ref{lem:MDS-PIRcapacity-achieving-matrix}.
The resulting achievable rate of Algorithm~\ref{alg:generation_QuerySet} after removing redundant $\tau$-sums according
to Lemma~\ref{lem:MDS-PIRcapacity-achieving-matrix} is given as
\begin{IEEEeqnarray*}{rCl}
  \label{eq:Cap_proof}
  \const{R} & \overset{(a)}{=} &\frac{k \nu^\mu}{n\sum_{\tau=1}^{\mu}
    \Bigl({\mu\choose\tau}-{\mu-r\choose\tau}\Bigr)\kappa^{\mu-(\tau-1)}(\nu-\kappa)^{\tau-1}}
  \\
  & \overset{(b)}{=} &\frac{\kappa\nu^\mu}{\nu\sum_{\tau=1}^{\mu}
    \Bigl({\mu\choose\tau}-{\mu-r\choose\tau}\Bigr)\kappa^{\mu-(\tau-1)}(\nu-\kappa)^{\tau-1}}
  \\
  & = &\frac{\nu^\mu\bigl(\frac{\nu-\kappa}{\nu}\bigr)}{\sum_{\tau=1}^{\mu}
    \Bigl({\mu\choose\tau}-{\mu-r\choose\tau}\Bigr)\kappa^{\mu-\tau}(\nu-\kappa)^\tau}
  \\
  &  & \qquad\quad \vdots
  \\[2mm]
  & \overset{(c)}{=}  &\frac{\nu^\mu\big(1-\frac{\kappa}{\nu}\big)}{\nu^\mu\!-\kappa^r \nu^{\mu-r}}
  \\[1mm]
  & = &\Bigl(1-\frac{\kappa}{\nu}\Bigr)\inv{\left[1-\Bigl(\frac{\kappa}{\nu}\Bigr)^r\right]},
\end{IEEEeqnarray*}
where we define $\binom{m}{n}\triangleq  0$ if $m<n$; (a) follows from the PLC rate in Definition~\ref{def:def_PLCrate}, the fact that $Q_j^{(v)}(\set{D};\tau) \cup Q_j^{(v)}(\set{U};\tau)$ contains $\kappa^{\mu-(\tau-1)}(\nu-\kappa)^{\tau-1}$ $\tau$-sums for every $\tau$-sum type, and
Lemma~\ref{lem:MDS-PIRcapacity-achieving-matrix}; (b) follows from Definition~\ref{def:MDS-PIRcapacity-achieving-codes};
and (c) follows by adapting similar steps as in the proof given in \cite{ObeadKliewer18_1}.

%%%%%%%%%%%%%%%%%%%%%%%%%%%%%%%%%%%%%%%%%%%%%%%%%%%%%%%%%%%%%%%%%%%%%%%%%%%%%%%%%%%%%%%%%%%%%%%%%%%%%%%%%%%%%%%%%%%%%%%%%%
%\vspace{-0.5ex} 
\section{Converse Proof}
\label{sec:converse-proof_coded-PLC}
In \cite{LinKumarRosnesGraellAmat18_2app,LinKumarRosnesGraellAmat18_3sub}, the PIR capacity of MDS-PIR capacity-achieving codes is shown to be equal to
the MDS-PIR capacity. In this section, we adapt the converse proof of \cite[Thm.~4]{LinKumarRosnesGraellAmat18_3sub} to
the scenario of the linearly-coded PLC problem. We will show that the PLC capacity of MDS-PIR capacity-achieving codes
is equal to  \eqref{eq:PLCcapacity}. Before we proceed with the converse proof, we give some
general results that are useful for the proof.
% in the converse proof.  % without specifying any given PLC protocol can be described as follows.

\begin{enumerate}
  % \item Assume that the user asks for the $v$-th linear combination, $v\in [1:\mu]$, from the $\mu$ candidate linear
  %   functions. Given a query $Q_j^{(v)}$ sent to the $j$-th database, the answer $A_j^{(v)}$ received by the user is a
  %   function of $Q_j^{(v)}$ and the $f$ coded chunks (denoted by
  %   $\vect{C}_j\eqdef\trans{\bigl(C^{(1)}_{1,j},\ldots,C^{(1)}_{\beta,j},C^{(2)}_{1,j},\ldots,C^{(f)}_{\beta,j}\bigr)}$)
  %   that are stored in the $j$-th database. Therefore,
  % \item Denote by
  %   $\vect{C}_j\eqdef\trans{\bigl(C^{(1)}_{1,j},\ldots,C^{(1)}_{\beta,j},C^{(2)}_{1,j},\ldots,C^{(f)}_{\beta,j}\bigr)}$
  %   the $f$ coded chunks that are stored in the $j$-th database. Then,
  %   \begin{IEEEeqnarray}{rCL}
  %     \bigHPcond{A_j^{(v)}}{Q_j^{(v)},\vmat{W}^{[1:f]}} & =
  %     &\bigHPcond{A_j^{(v)}}{Q_j^{(v)},\vect{C}_j}=0.\IEEEeqnarraynumspace\label{eq:answers}
  %   \end{IEEEeqnarray}
  % \item From the condition of privacy, the $j$-th database should not be able to differentiate between the answers
  %   $A_j^{(v)}$ and $A_j^{(v')}$ when the user requests $\vmat{X}^{(v)}$, $v\neq v'$, $v,v'\in [1:\mu]$. Hence,
\item From the condition of privacy,
  \begin{IEEEeqnarray}{rCl}
    \bigHPcond{A_j^{(v)}}{\vmat{X}^{(v)},\set{Q}}=\bigHPcond{A_j^{(v')}}{\vmat{X}^{(v)},\set{Q}},
    \label{eq:indistinct-answers}
  \end{IEEEeqnarray}
  where $v\neq v'$, $v,v'\in [1:\mu]$, and 
  $\set{Q}\eqdef\bigl\{Q^{(v)}_j\colon v\in[1:\mu], j\in[1:n]\bigr\}$ denotes the set of all possible queries  made
  by the user. 
  Although this seems to be intuitively true, a proof of this property 
  is still required and can be found
  in \cite[Lem.~3]{XuZhang17_1sub}.

\item Consider a PLC protocol for a coded DSS that uses an $[n,k]$ code $\code{C}$ to store $f$ messages. For any subset
  of linear combinations $\set{V}\subseteq[1:\mu]$ and for any information set $\set{I}$ of $\code{C}$, we have
  \begin{IEEEeqnarray}{rCl}
    \bigHPcond{A^{(v)}_\set{I}}{\vmat{X}^{\set{V}},\set{Q}}& = &
    \sum_{j\in\set{I}}\bigHPcond{A^{(v)}_j}{\vmat{X}^{\set{V}},\set{Q}}.
    \label{eq:independent_kAnswers}
  \end{IEEEeqnarray}
  The proof uses the linear independence of the columns of a generator matrix of $\code{C}$ corresponding to an
  information set, and can be seen as a simple extension of \cite[Lem.~2]{BanawanUlukus18_1} or
  \cite[Lem.~4]{XuZhang17_1sub}.
  This argument applies to the case of PLC due to the fact that $A^{(v)}_\set{I}$ is
  still a deterministic function of  independent random variables $\{\vect{C}_j\colon j \in \set{I} \}$ and $\set{Q}$, where $\vect{C}_j\eqdef\trans{\bigl(C^{(1)}_{1,j},\ldots,C^{(1)}_{\beta,j},C^{(2)}_{1,j},\ldots,C^{(f)}_{\beta,j}\bigr)}$ denotes 
  the $f$ coded chunks that are stored in the $j$-th database.
\end{enumerate}

Next, we state Shearer's Lemma, which represents a very useful entropy method for combinatorial problems.
\begin{lemma}[Shearer's Lemma \cite{Radhakrishnan03_1}]
  \label{lem:Shearer-lemma}
  Let $\collect{S}$ be a collection of subsets of $[1:n]$, with each $j\in[1:n]$ included in at least $\kappa$ members
  of $\collect{S}$. For random variables $Z_1,\ldots,Z_n$, we have
  \begin{IEEEeqnarray*}{rCl}
    \sum_{\set{S}\in\collect{S}}\eHP{Z_\set{S}}\geq \kappa\eHP{Z_1,\ldots,Z_n}.
    %\label{eq:Shearer-lemma}
  \end{IEEEeqnarray*}
\end{lemma}

For our converse proof for the coded PLC problem, we also need the following lemma, whose proof has been omitted due to
lack of space.
\begin{lemma}
  \label{lem:uniform_dist}
  Consider the linear mapping $\mat{V}=(v_{i,j})$ defined in \eqref{eq:linear-mappingV} with $\rank{\mat{V}}=r$ where
  $v_{i_1,j_1},\ldots,v_{i_r,j_r}$ are the entries corresponding to the pivot elements of $\mat{V}$. It follows that
  $\bigl(\vmat{X}^{(i_1)},\ldots,\vmat{X}^{(i_h)}\bigr)$ and $\bigl(\vmat{W}^{(j_1)},\ldots,\vmat{W}^{(j_h)}\bigr)$ are
  identically distributed, for some $h\in [1:r]$. In other words,
  $\bigHP{\vmat{X}^{(i_1)},\ldots,\vmat{X}^{(i_h)}}=h\const{L}$, $h\in [1:r]$.
\end{lemma}
%\begin{IEEEproof}
%  See Appendix~\ref{sec:proof_uniform-dist}.
%\end{IEEEproof}

Now, we are ready for the converse proof. By \cite[Lem.~2]{LinKumarRosnesGraellAmat18_1}, since the code $\code{C}$ is
MDS-PIR capacity-achieving, there exist $\nu$ information sets $\set{I}_1,\ldots,\set{I}_\nu$ such that each coordinate
$j\in[1:n]$ is included in exactly $\kappa$ members of $\collect{I}=\{\set{I}_1,\ldots,\set{I}_\nu\}$ with
$\frac{\kappa}{\nu}=\frac{k}{n}$.

Applying the chain rule of entropy we have
\begin{IEEEeqnarray*}{rCl}
  \bigHPcond{A^{(v)}_{[1:n]}}{\vmat{X}^\set{V},\set{Q}}\geq\bigHPcond{A^{(v)}_{\set{I}_i}}{\vmat{X}^\set{V},\set{Q}},
  \quad\forall\,i\in [1:\nu].
\end{IEEEeqnarray*}
  
Let $v\in\set{V}$ and $v'\in\cset{\set{V}}\eqdef[1:\mu]\setminus\set{V}$. Following similar steps as in the proof given
in \cite{BanawanUlukus18_1,XuZhang17_1sub}, we get
\begin{IEEEeqnarray}{rCl}
  \IEEEeqnarraymulticol{3}{l}{%
    \nu\bigHPcond{A^{(v)}_{[1:n]}}{\vmat{X}^\set{V},\set{Q}} }\nonumber\\*\quad%
  &\geq &\sum_{i=1}^\nu\bigHPcond{A^{(v)}_{\set{I}_i}}{\vmat{X}^\set{V},\set{Q}}
  \nonumber\\
  & = &\sum_{i=1}^\nu\left(\sum_{j\in\set{I}_i}\bigHPcond{A^{(v)}_j}{\vmat{X}^\set{V},\set{Q}}\right)
  \label{eq:use_kAnswers1}\\[1mm]
  & = &\sum_{i=1}^\nu\left(\sum_{j\in\set{I}_i}\bigHPcond{A^{(v')}_j}{\vmat{X}^\set{V},\set{Q}}\right)
  \label{eq:use_indistinct-answers}\\
  & = &\sum_{i=1}^\nu\bigHPcond{A^{(v')}_{\set{I}_i}}{\vmat{X}^\set{V},\set{Q}}
  \label{eq:use_kAnswers2}\\[1mm]
  & \geq &\kappa\bigHPcond{A^{(v')}_{[1:n]}}{\vmat{X}^\set{V},\set{Q}}
  \label{eq:use_Shearer-lemma}\\[1mm]
  & = & \kappa\Bigl[\bigHPcond{A^{(v')}_{[1:n]},\vmat{X}^{(v')}}{\vmat{X}^\set{V},\set{Q}}
  \nonumber\\
  && \qquad-\bigHPcond{\vmat{X}^{(v')}}{A^{(v')}_{[1:n]},\vmat{X}^\set{V},\set{Q}}\Bigr]
  \nonumber\\
  & = &\kappa\Bigl[\bigHPcond{\vmat{X}^{(v')}}{\vmat{X}^\set{V},\set{Q}}
  \nonumber\\
  && \qquad+\>\bigHPcond{A^{(v')}_{[1:n]}}{\vmat{X}^\set{V},\vmat{X}^{(v')},\set{Q}}-0\Bigr]
  \label{eq:use_recovery1}\\
  & = & \kappa\Bigl[\bigHPcond{\vmat{X}^{(v')}}{\vmat{X}^\set{V}}
  \!+\!\bigHPcond{A^{(v')}_{[1:n]}}{\vmat{X}^\set{V},\vmat{X}^{(v')},\set{Q}}\Bigr],\label{eq:use_independence}
\end{IEEEeqnarray}
where \eqref{eq:use_kAnswers1} and \eqref{eq:use_kAnswers2} follow from \eqref{eq:independent_kAnswers};
\eqref{eq:use_indistinct-answers} is because of \eqref{eq:indistinct-answers}; \eqref{eq:use_Shearer-lemma} is due to
Shearer's Lemma; \eqref{eq:use_recovery1} is from the fact that the $v'$-th linear combination $\vmat{X}^{(v')}$ is
determined by the answers $A^{(v')}_{[1:n]}$ and all possible queries $\set{Q}$; and finally, \eqref{eq:use_independence}
follows from the independence between all possible queries and the messages. Therefore, we can conclude that
\begin{IEEEeqnarray}{rCl}
  \bigHPcond{&A^{(v)}_{[1:n]}&}{\vmat{X}^\set{V},\set{Q}}\nonumber\\
  & \geq & \frac{\kappa}{\nu}\bigHPcond{\vmat{X}^{(v')}}{\vmat{X}^\set{V}}+\frac{\kappa}{\nu}
  \bigHPcond{A^{(v')}_{[1:n]}}{\vmat{X}^\set{V},\vmat{X}^{(v')},\set{Q}}
  \nonumber\\
  & = & \frac{k}{n}
  \bigHPcond{\vmat{X}^{(v')}}{\vmat{X}^\set{V}}
  +\frac{k}{n}\bigHPcond{A^{(v')}_{[1:n]}}{\vmat{X}^\set{V},\vmat{X}^{(v')},\set{Q}},
  \IEEEeqnarraynumspace\label{eq:use_MDS-PIRcapaciy-achieving-codes}
\end{IEEEeqnarray}
where we have used Definition~\ref{def:MDS-PIRcapacity-achieving-codes} to obtain
\eqref{eq:use_MDS-PIRcapaciy-achieving-codes}.

Since there are in total $\mu$ linear combinations and $\set{L}\eqdef\{\ell_1,\ldots,\ell_r\}\subseteq [1:\mu]$ is the
set of row indices corresponding to the selected basis for the row space of $\mat{V}$, we can recursively use
\eqref{eq:use_MDS-PIRcapaciy-achieving-codes} $r-1$ times to obtain
\begin{IEEEeqnarray}{rCl}    
  \IEEEeqnarraymulticol{3}{l}{%
    \bigHPcond{A_{[1:n]}^{(\ell_1)}}{\vmat{X}^{(\ell_1)},\set{Q}} }\nonumber\\*\quad%
  & \geq &\sum_{v=1}^{r-1}\Bigl(\frac{k}{n}\Bigr)^v
  \bigHPcond{\vmat{X}^{(\ell_{v+1})}}{\vmat{X}^{\{\ell_1,\ldots,\ell_{v}\}}}
  \nonumber\\
  && \qquad+\Bigl(\frac{k}{n}\Bigr)^{r-1}\bigHPcond{A_{[1:n]}^{(\ell_r)}}{\vmat{X}^{\{\ell_1,\ldots,\ell_r\}},\set{Q}}
  \nonumber\\
  & \geq &\sum_{v=1}^{r-1}\Bigl(\frac{k}{n}\Bigr)^v
  \bigHPcond{\vmat{X}^{(\ell_{v+1})}}{\vmat{X}^{\{\ell_1,\ldots,\ell_{v}\}}}
  \label{eq:nonegative-entropy}\\
  & = &\sum_{v=1}^{r-1}\Bigl(\frac{k}{n}\Bigr)^v\const{L},\label{eq:use_rank-lemma}
\end{IEEEeqnarray}
where \eqref{eq:nonegative-entropy} follows from the nonnegativity of entropy.
% from \eqref{eq:answers} and the fact that there exist $r$ messages
% $\vmat{W}^{\{j_1,\ldots,j_r\}}\subseteq\vmat{W}^{[1:f]}$, that are linearly mapped from
% $\vmat{X}^{\{\ell_1,\ldots,\ell_r\}}$.
\eqref{eq:use_rank-lemma} holds since it follows from Lemma~\ref{lem:uniform_dist} that
$\bigHPcond{\vmat{X}^{(\ell_{v+1})}}{\vmat{X}^{\{\ell_1,\ldots,\ell_{v}\}}}=\bigHP{\vmat{X}^{(\ell_{v+1})}}=
\const{L}$. Here, we also remark that the recursive steps follow the same principle of the general converse for
dependent PIR (DPIR) from \cite[Thm.~1]{ChenWangJafar18_1}. In \cite{ChenWangJafar18_1}, the authors claim that the
general converse for the DPIR problem strongly depends on the chosen permutation of the indices of the candidate
functions. However, the best index permutation for the candidate linear functions for the PLC problem results from
finding a basis for $\mat{V}$. Now,
\begin{IEEEeqnarray}{rCl}
  \const{L}& = &\HP{\vmat{X}^{(\ell_1)}}
  \nonumber\\
  & = &\bigHPcond{\vmat{X}^{(\ell_1)}}{\set{Q}}
  -\underbrace{\bigHPcond{\vmat{X}^{(\ell_1)}}{A^{(\ell_1)}_{[1:n]},\set{Q}}}_{=0}
  \label{eq:use_recovery2}\\
  & = &\bigMIcond{\vmat{X}^{(\ell_1)}}{A^{(\ell_1)}_{[1:n]}}{\set{Q}}
  \nonumber\\[1mm]
  & = &\BigHPcond{A^{(\ell_1)}_{[1:n]}}{\set{Q}}-\BigHPcond{A^{(\ell_1)}_{[1:n]}}{\vmat{X}^{(\ell_1)},\set{Q}}
  \nonumber\\
  & \leq &\BigHPcond{A^{(\ell_1)}_{[1:n]}}{\set{Q}}-\sum_{v=1}^{r-1}\Bigl(\frac{k}{n}\Bigr)^v\const{L},
  \label{eq:use_recursive-step}
\end{IEEEeqnarray}
where \eqref{eq:use_recovery2} follows since any message is independent of the queries $\set{Q}$, and knowing the
answers $A^{(\ell_1)}_{[1:n]}$ and the queries $\set{Q}$, one can determine $\vmat{X}^{(\ell_1)}$;
\eqref{eq:use_recursive-step} holds because of \eqref{eq:use_rank-lemma}.

Finally, the converse proof is completed by showing that
\begin{IEEEeqnarray}{rCl}
  \const{R}& = &\frac{\const{L}}{\sum_{j=1}^n \BigHP{A^{(\ell_1)}_j}}
  \nonumber\\
  & \leq &\frac{\const{L}}{\BigHP{A^{(\ell_1)}_{[1:n]}}}
  \label{eq:use_chain-rule}\\
  & \leq &\frac{\const{L}}{\BigHPcond{A^{(\ell_1)}_{[1:n]}}{\set{Q}}}
  \label{eq:use_conditioning-entropy}\\
  & \leq & \frac{1}{1+\sum_{v=1}^{r-1}\bigl(\frac{k}{n}\bigr)^{v}}=\const{C}_{\textnormal{PLC}},\label{eq:14}
\end{IEEEeqnarray}
where \eqref{eq:use_chain-rule} holds because of the chain rule of entropy, \eqref{eq:use_conditioning-entropy} is due to
the fact that conditioning reduces entropy, and we apply \eqref{eq:use_recursive-step} to obtain \eqref{eq:14}.

%\vspace{-0.5ex} 
\section{Conclusion}
\label{sec:conclusion}
We have provided the capacity of PLC for DSSs, where data is encoded and stored using an
arbitrary linear code from the class of MDS-PIR capacity-achieving codes. Interestingly, the capacity is equal to the
corresponding PIR capacity. Thus, privately retrieving arbitrary linear combinations of the stored messages does not
incur any overhead in rate compared to retrieving a single message from the databases.

\balance

%%%%%%%%%%%%%%%%%%%%%%%%%%%%%%%%%%%%%%%%%%%%%%%%%%%%%%%%%%%%%%%%%%%%%%%%%%%%% 
% trigger a \newpage just before the given reference
% number - used to balance the columns on the last page
% adjust value as needed - may need to be readjusted if
% the document is modified later
% \IEEEtriggeratref{3}
% The "triggered" command can be changed if desired:
%\IEEEtriggercmd{\enlargethispage{-5in}}

% references section
% \balance

%%%%%%%%%%%%%%%%%%%%%%%%%%%%%%%%%%%%%%%%%%%%%%%%%%%%%%%%%%%%%%%%%%%%%%%%%%%%% 
\bibliographystyle{IEEEtran} 
\bibliography{./defshort1,./biblio1,./references}

% Generated by IEEEtran.bst, version: 1.14 (2015/08/26)
\begin{thebibliography}{10}
\providecommand{\url}[1]{#1}
\csname url@samestyle\endcsname
\providecommand{\newblock}{\relax}
\providecommand{\bibinfo}[2]{#2}
\providecommand{\BIBentrySTDinterwordspacing}{\spaceskip=0pt\relax}
\providecommand{\BIBentryALTinterwordstretchfactor}{4}
\providecommand{\BIBentryALTinterwordspacing}{\spaceskip=\fontdimen2\font plus
\BIBentryALTinterwordstretchfactor\fontdimen3\font minus
  \fontdimen4\font\relax}
\providecommand{\BIBforeignlanguage}[2]{{%
\expandafter\ifx\csname l@#1\endcsname\relax
\typeout{** WARNING: IEEEtran.bst: No hyphenation pattern has been}%
\typeout{** loaded for the language `#1'. Using the pattern for}%
\typeout{** the default language instead.}%
\else
\language=\csname l@#1\endcsname
\fi
#2}}
\providecommand{\BIBdecl}{\relax}
\BIBdecl

\bibitem{ChorKushilevitzGoldreichSudan98_1}
B.~Chor, E.~Kushilevitz, O.~Goldreich, and M.~Sudan, ``Private information
  retrieval,'' \emph{J. ACM}, vol.~45, no.~6, pp. 965--981, Nov. 1998.

\bibitem{Yekhanin10_1}
S.~Yekhanin, ``Private information retrieval,'' \emph{Commun. ACM}, vol.~53,
  no.~4, pp. 68--73, Apr. 2010.

\bibitem{ShahRashmiRamchandran14_1}
N.~B. Shah, K.~V. Rashmi, and K.~Ramchandran, ``One extra bit of download
  ensures perfectly private information retrieval,'' in \emph{Proc. IEEE Int.
  Symp. Inf. Theory}, Honolulu, HI, USA, Jun. 29 -- Jul. 4, 2014, pp. 856--860.

\bibitem{ChanHoYamamoto15_1}
T.~H. Chan, S.-W. Ho, and H.~Yamamoto, ``Private information retrieval for
  coded storage,'' in \emph{Proc. IEEE Int. Symp. Inf. Theory}, Hong Kong,
  China, Jun. 14--19, 2015, pp. 2842--2846.

\bibitem{TajeddineGnilkeElRouayheb18_1app}
R.~Tajeddine, O.~W. Gnilke, and S.~El~Rouayheb, ``Private information retrieval
  from {MDS} coded data in distributed storage systems,'' 2018, to app. in
  \itshape IEEE Trans. Inf. Theory\upshape.

\bibitem{SunJafar17_1}
H.~Sun and S.~A. Jafar, ``The capacity of private information retrieval,''
  \emph{IEEE Trans. Inf. Theory}, vol.~63, no.~7, pp. 4075--4088, Jul. 2017.

\bibitem{SunJafar18_2}
------, ``The capacity of robust private information retrieval with colluding
  databases,'' \emph{IEEE Trans. Inf. Theory}, vol.~64, no.~4, pp. 2361--2370,
  Apr. 2018.

\bibitem{BanawanUlukus18_1}
K.~Banawan and S.~Ulukus, ``The capacity of private information retrieval from
  coded databases,'' \emph{IEEE Trans. Inf. Theory}, vol.~64, no.~3, pp.
  1945--1956, Mar. 2018.

\bibitem{FreijHollantiGnilkeHollantiKarpuk17_1}
R.~Freij-Hollanti, O.~W. Gnilke, C.~Hollanti, and D.~A. Karpuk, ``Private
  information retrieval from coded databases with colluding servers,''
  \emph{SIAM J. Appl. Algebra Geom.}, vol.~1, no.~1, pp. 647--664, Nov. 2017.

\bibitem{KumarRosnesGraellAmat17_1}
S.~Kumar, E.~Rosnes, and A.~Graell~i Amat, ``Private information retrieval in
  distributed storage systems using an arbitrary linear code,'' in \emph{Proc.
  IEEE Int. Symp. Inf. Theory}, Aachen, Germany, Jun. 25--30, 2017, pp.
  1421--1425.

\bibitem{MirmohseniMaddahAli17_1sub}
\BIBentryALTinterwordspacing
M.~Mirmohseni and M.~A. Maddah-Ali, ``Private function retrieval,'' Nov. 2017,
  arXiv:1711.04677v2 [cs.IT]. [Online]. Available:
  \url{https://arxiv.org/abs/1711.04677}
\BIBentrySTDinterwordspacing

\bibitem{SunJafar17_1sub}
\BIBentryALTinterwordspacing
H.~Sun and S.~A. Jafar, ``The capacity of private computation,'' Oct. 2017,
  arXiv:1710.11098v3 [cs.IT]. [Online]. Available:
  \url{https://arxiv.org/abs/1710.11098}
\BIBentrySTDinterwordspacing

\bibitem{Karpuk18_1}
D.~Karpuk, ``Private computation of systematically encoded data with colluding
  servers,'' in \emph{Proc. IEEE Int. Symp. Inf. Theory}, Vail, CO, USA, Jun.
  17--22, 2018, pp. 2112--2116.

\bibitem{ObeadKliewer18_1}
S.~A. Obead and J.~Kliewer, ``Achievable rate of private function retrieval
  from {MDS} coded databases,'' in \emph{Proc. IEEE Int. Symp. Inf. Theory},
  Vail, CO, USA, Jun. 17--22, 2018, pp. 2117--2121.

\bibitem{ChenWangJafar18_1}
Z.~Chen, Z.~Wang, and S.~Jafar, ``The asymptotic capacity of private search,''
  in \emph{Proc. IEEE Int. Symp. Inf. Theory}, Vail, CO, USA, Jun. 17--22,
  2018, pp. 2122--2126.

\bibitem{KumarLinRosnesGraellAmat17_1sub}
\BIBentryALTinterwordspacing
S.~Kumar, H.-Y. Lin, E.~Rosnes, and A.~Graell~i Amat, ``Achieving maximum
  distance separable private information retrieval capacity with linear
  codes,'' Dec. 2017, arXiv:1712.03898v3 [cs.IT]. [Online]. Available:
  \url{https://arxiv.org/abs/1712.03898}
\BIBentrySTDinterwordspacing

\bibitem{LinKumarRosnesGraellAmat18_1}
H.-Y. Lin, S.~Kumar, E.~Rosnes, and A.~Graell~i Amat, ``An {MDS}-{PIR}
  capacity-achieving protocol for distributed storage using non-{MDS} linear
  codes,'' in \emph{Proc. IEEE Int. Symp. Inf. Theory}, Vail, CO, USA, Jun.
  17--22, 2018, pp. 966--970.

\bibitem{LinKumarRosnesGraellAmat18_2app}
\BIBentryALTinterwordspacing
------, ``Asymmetry helps: Improved private information retrieval protocols for
  distributed storage,'' to be pres. at \itshape IEEE Inf. Theory
  Workshop\upshape, Guangzhou, China, Nov. 25--29, 2018. [Online]. Available:
  \url{https://arxiv.org/abs/1806.01342}
\BIBentrySTDinterwordspacing

\bibitem{LinKumarRosnesGraellAmat18_3sub}
\BIBentryALTinterwordspacing
------, ``On the fundamental limit of private information retrieval for coded
  distributed storage,'' Aug. 2018, arXiv:1808.09018v1 [cs.IT]. [Online].
  Available: \url{https://arxiv.org/abs/1808.09018}
\BIBentrySTDinterwordspacing

\bibitem{FreijHollantiGnilkeHollanti17_1sub}
\BIBentryALTinterwordspacing
R.~Freij-Hollanti, O.~W. Gnilke, C.~Hollanti, A.-L. Horlemann-Trautmann,
  D.~Karpuk, and I.~Kubjas, ``$t$-private information retrieval schemes using
  transitive codes,'' Dec. 2017, arXiv:1712.02850v1 [cs.IT]. [Online].
  Available: \url{https://arxiv.org/abs/1712.02850}
\BIBentrySTDinterwordspacing

\bibitem{XuZhang17_1sub}
\BIBentryALTinterwordspacing
J.~Xu and Z.~Zhang, ``On sub-packetization of capacity-achieving {PIR} schemes
  for {MDS} coded databases,'' Dec. 2017, arXiv:1712.02466v2 [cs.IT]. [Online].
  Available: \url{http://arxiv.org/abs/1712.02466}
\BIBentrySTDinterwordspacing

\bibitem{Radhakrishnan03_1}
J.~Radhakrishnan, ``Entropy and counting,'' in \emph{Computational mathematics,
  modelling and algorithms}, J.~C. Misra, Ed.\hskip 1em plus 0.5em minus
  0.4em\relax Narosa Publishing House, 2003, pp. 146--168.

\end{thebibliography}

\end{document}